\documentclass[aps,pre,reprint,floatfix,superscriptaddress,amsfonts,amsmath,amssymb,citeautoscript]{revtex4-2}
\usepackage[pdftex]{graphicx}
\usepackage{color}
\usepackage[noEnd=false,indLines=false]{algpseudocodex}

\begin{document}

\title{Algorithm to generate hierarchical structure of desiccation crack patterns}

\author{Yuri Yu. Tarasevich}
\email[Corresponding author: ]{tarasevich@asu-edu.ru}
\affiliation{Laboratory of Mathematical Modeling, Astrakhan Tatishchev State University, Astrakhan, Russia}

\author{Andrei V. Eserkepov}
\email{dantealigjery49@gmail.com}
\affiliation{Laboratory of Mathematical Modeling, Astrakhan Tatishchev State University, Astrakhan, Russia}

\author{Irina V. Vodolazskaya}
\email{vodolazskaya\_agu@mail.ru}
\affiliation{Laboratory of Mathematical Modeling, Astrakhan Tatishchev State University, Astrakhan, Russia}

\date{\today}

\begin{abstract}
We propose an algorithm generating planar networks which structure resembles a hierarchical structure of desiccation crack patterns.

\end{abstract}

\maketitle

\section{Introduction: Hierarchical structure of cracks arising during drying of thin films of colloids and polymers\label{sec:intro}}

When thin films of colloids dry, a hierarchical structure of cracks arises~\cite{Bohn2005,Bohn2005a,Perna2011,Tang2011,Kumar2021,Voronin2022}. In particular, a newly forming crack connects to an earlier crack at an angle close to $90^\circ$~\cite{Bohn2005}; according to the classification~\cite{Gray1976}, such crack connections are of the $T$-type. If, after digitizing the image and transforming it into a graph embedded in the plane, the angle between adjacent edges is close to $180^\circ$, then such edges are considered to be parts of the same crack. According to~\cite{Bohn2005}, primary cracks do not connect to any other cracks; their ends are outside the observation window. Cracks that terminate on primary cracks are called secondary cracks. In general, a crack of order $n$ terminates at least at one of its ends at a crack of $n-1$-th order. In addition, nuclei (defects) may form from which cracks begin to grow in the form of a three-pointed star with angles between cracks of approximately $120^\circ$; according to the classification~\cite{Gray1976}, such crack junctions belong to the $Y$-type. According to~\cite{Gray1976}, $X$-shaped crack junctions are common in natural crack patterns, but higher-order junctions are generally absent. It can be assumed that $X$-shaped crack junctions are a degenerate case of two $T$-shaped cracks, when the edge is very short, or it is a four-pointed star extending from the nucleus. In~\cite{Kumar2021}, an alternative classification based on crack width analysis was proposed. With this classification, it turns out that parts of the same crack according to the classification~\cite{Bohn2005} belong to different generations. According to~\cite{Kumar2021}, there is a pattern between the width and total length of cracks of different generations. In addition, only $T$-shaped and $Y$-shaped cracks were observed in the work~\cite{Kumar2021}.

The resulting cracks can be filled with a conductive material (Ag, Cu, Ni, etc.), which leads to the creation of a random conductive network (transparent electrode, transparent conductive film)~\cite{Voronin2023}. Knowledge of the hierarchical structure of cracks is important, in particular, for calculating electrical conductivity, since cracks of different orders have different thicknesses~\cite{Kumar2021}: the higher the order, the thinner the crack. The crack width distributions given in~\cite{Kumar2021} allow us to roughly estimate the ratio of the thickness of third-, second-, and first-order cracks as $0.6:0.8:1$.

\section{Recursive algorithm simulating the hierarchical structure of cracks that occur during drying of thin films of colloids and polymers}

Various algorithms for simulating the structure of cracks formed in various materials are described in literature~\cite{Pons2010,Khatun2012,Roy2022,Haque2023,Leon2023,Noguchi2024,Liu2024}. The diversity of observed structures leads to a diversity of algorithms.

In developing our own algorithm, we aimed to ensure that the algorithm reproduces the hierarchical structure of cracks (both temporal and spatial hierarchy). However, we did not aim to reproduce in detail the geometric properties of real crack networks, in which cracks are often curved and enter each other almost perpendicularly. We took as a basis the Voronoi tessellation, which generally correctly reproduces the morphology of crack networks~\cite{Tarasevich2023}, with the exception of the hierarchical structure. Our algorithm imitates the temporal hierarchy of crack formation: a small number of seeds are placed in a given area, after which Voronoi tessellation is performed (primary cracks). Assuming that each of the areas into which the system is divided becomes independent, the partitioning procedure is repeated for each resulting cell separately until a given number density of cracks is achieved.

\begin{algorithmic}[1]
\LComment{Number of consecutive partitions}
\State set global int iterations
\LComment{Change in the number density with new partition}
\State set global int multiplier
\Function{RecursiveVoronoi}{iteration, concentration, points, outerVertices}
	\If{iteration $\leqslant$ iterations}
		\State vor $\gets$ \Call{voronoi}{points, outerVertices}
\LComment {construction of a partition by given seeds within given domain}
		\State newIteration $\gets$ iteration + 1
		\State newConcentration $\gets$ concentration * multiplier
		\State newPoints $\gets$ \Call{gen}{newConcentration}
 \LComment {generating a list of seeds}
		\For {cell $\in$ vor.cells}
			\State localPoints $\gets$ \Call{PointsInThisCell}{newPoints, cell.vertices}
 \LComment {we select those seeds that fell into the cell}
		\State 	\Call{RecursiveVoronoi}{newIteration, newConcentration, localPoints, cell.vertices}
\EndFor
\EndIf
    \EndFunction
\end{algorithmic}

\begin{figure}[!htbp]
  \centering
  \includegraphics[width=\columnwidth]{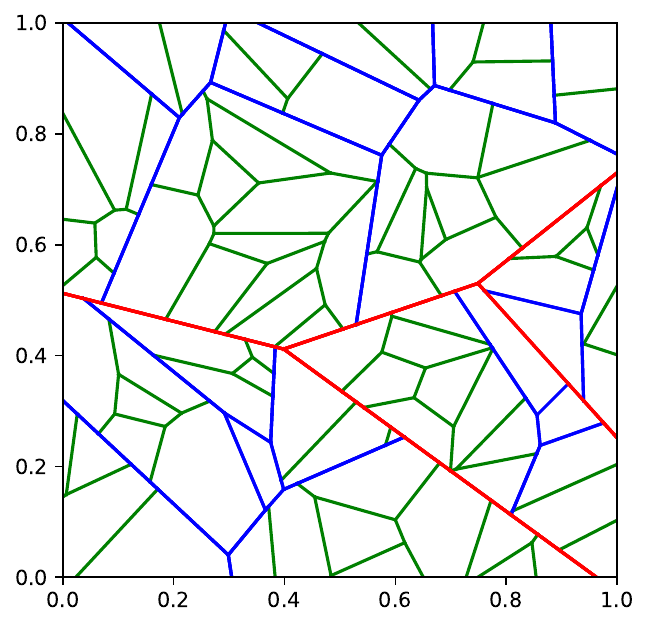}
  \caption{An example of a network obtained using a recursive algorithm.}\label{fig:PatternRecurs}
\end{figure}

Figure~\ref{fig:PatternRecurs} shows an example of a network obtained using a recursive algorithm. The edges of the cells obtained at the first iteration are shown in red, at the second in blue, and at the third in green. Note that the resulting hierarchy differs from the classification~\cite{Bohn2005}.

\subsection{Geometric and topological properties of networks obtained using a recursive algorithm}
\begin{figure}[!htbp]
  \centering
  \includegraphics[width=\columnwidth]{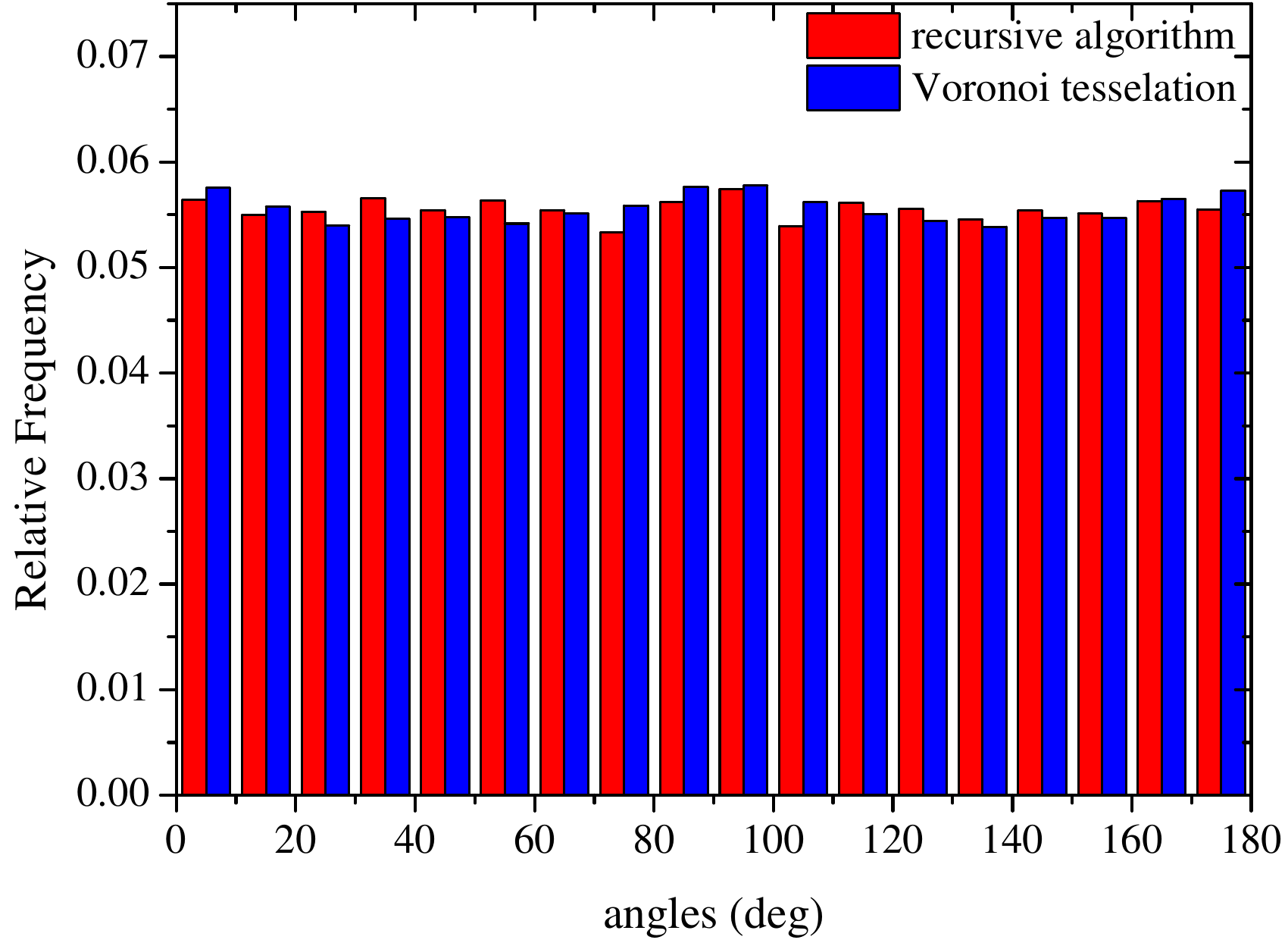}
  \caption{Distribution of cracks by orientations. Comparison of results obtained using the usual Voronoi tessellation and the recursive algorithm.}\label{fig:angles}
\end{figure}

Statistical analysis was performed using 1000 different networks obtained by a recursive algorithm. The regions used were $L \times L, L=1$ with the number density of seeds about 100.
Figure~\ref{fig:angles} shows that the distribution of edge orientations in the networks is close to equiprobable.

\begin{figure}[!htbp]
  \centering
  \includegraphics[width=\columnwidth]{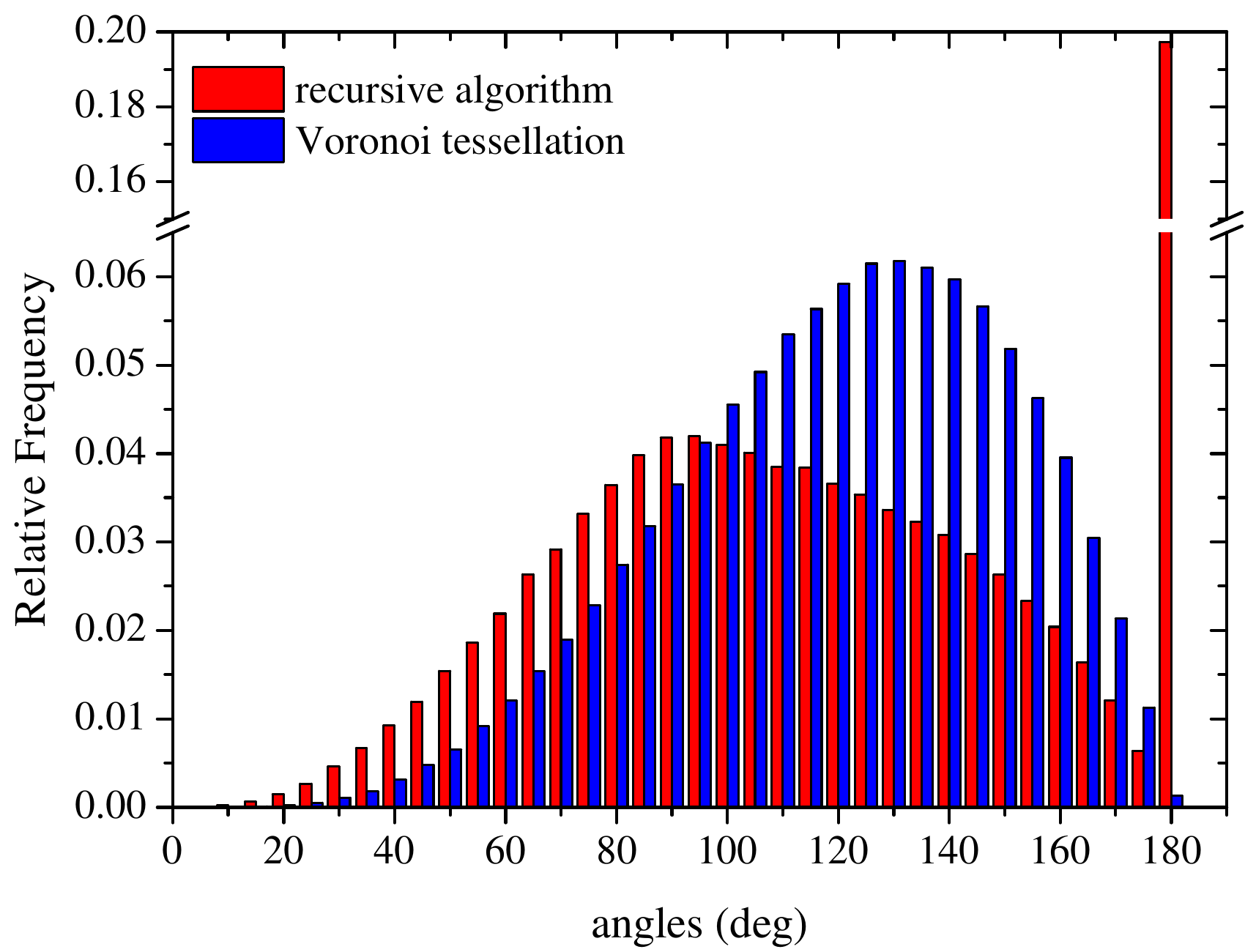}
  \caption{Distribution of angles between cracks. Comparison of results obtained using the usual Voronoi partition and the recursive algorithm.}\label{fig:CrackAngles}
\end{figure}

Figure~\ref{fig:CrackAngles} shows that the distribution of angles between the nearest edges in the networks is asymmetric, which is due to the superposition of two symmetric distributions: with the center $90^\circ$ and with the center $120^\circ$, in addition, there is a peak corresponding to the angles $180^\circ$. The distribution indicates the presence of $Y$- and $T$-cracks, according to the classification~\cite{Gray1976}. Note that in conventional Voronoi diagrams, the distribution is symmetric with a mode of about $120^\circ$ (all cracks are $Y$-shaped)~\cite{Tarasevich2023}. Using the recursive algorithm leads to the appearance of $T$-shaped cracks. The resulting distribution is close to that obtained when processing photographs of real samples~\cite{Akiba2017,Tarasevich2023} of crack patterns.

\begin{figure}[!htbp]
  \centering
  \includegraphics[width=\columnwidth]{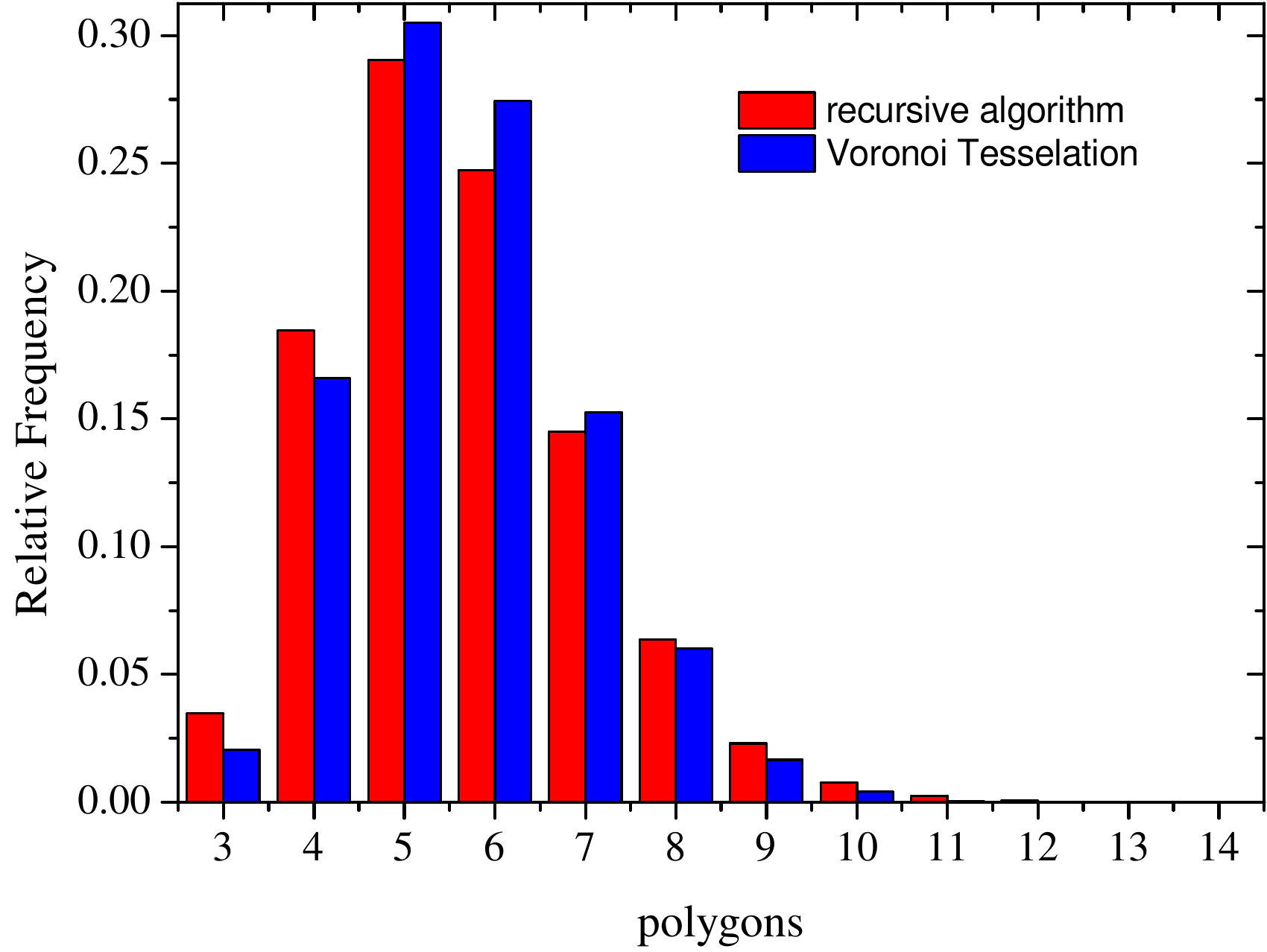}
  \caption{Distribution of cells by number of sides. Comparison of results obtained using the usual Voronoi partition and the recursive algorithm.}\label{fig:vertex}
\end{figure}

The figure~\ref{fig:vertex} shows the distribution of cells by the number of sides. The systems have the largest number of cells with 5 vertices, followed in descending order by hexagons, quadrangles, and heptagons. The obtained distribution is consistent with the data from the analysis of real fracture networks~\cite{Akiba2017}.

\begin{figure}[!htbp]
  \centering
  \includegraphics[width=\columnwidth]{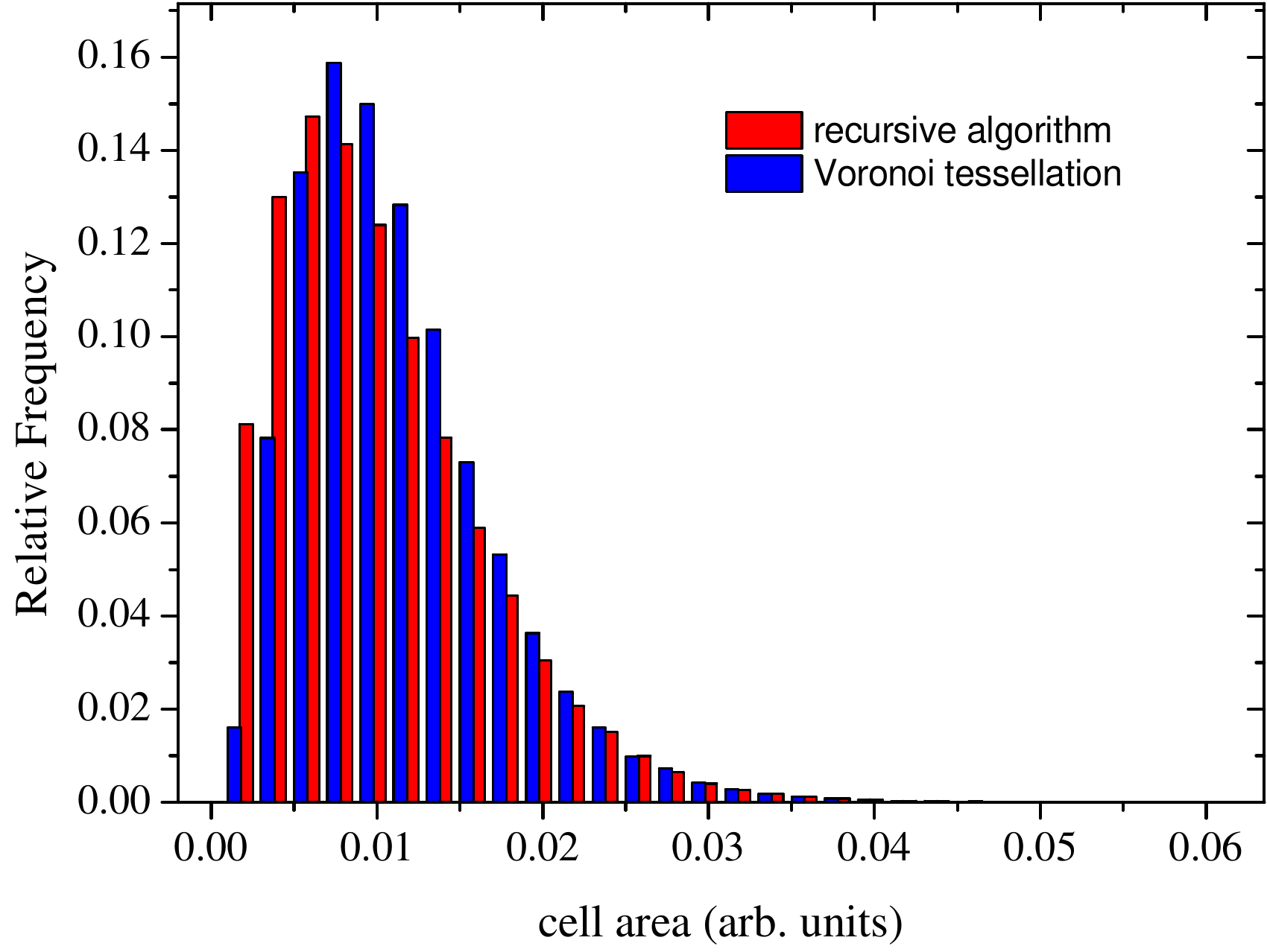}
  \caption{Distribution of cell sizes. Comparison of results obtained using the usual Voronoi tessellation and the recursive algorithm.}\label{fig:areas}
\end{figure}

Figure~\ref{fig:areas} shows the distribution of cell sizes.

\begin{figure}[!htbp]
  \centering
  \includegraphics[width=\columnwidth]{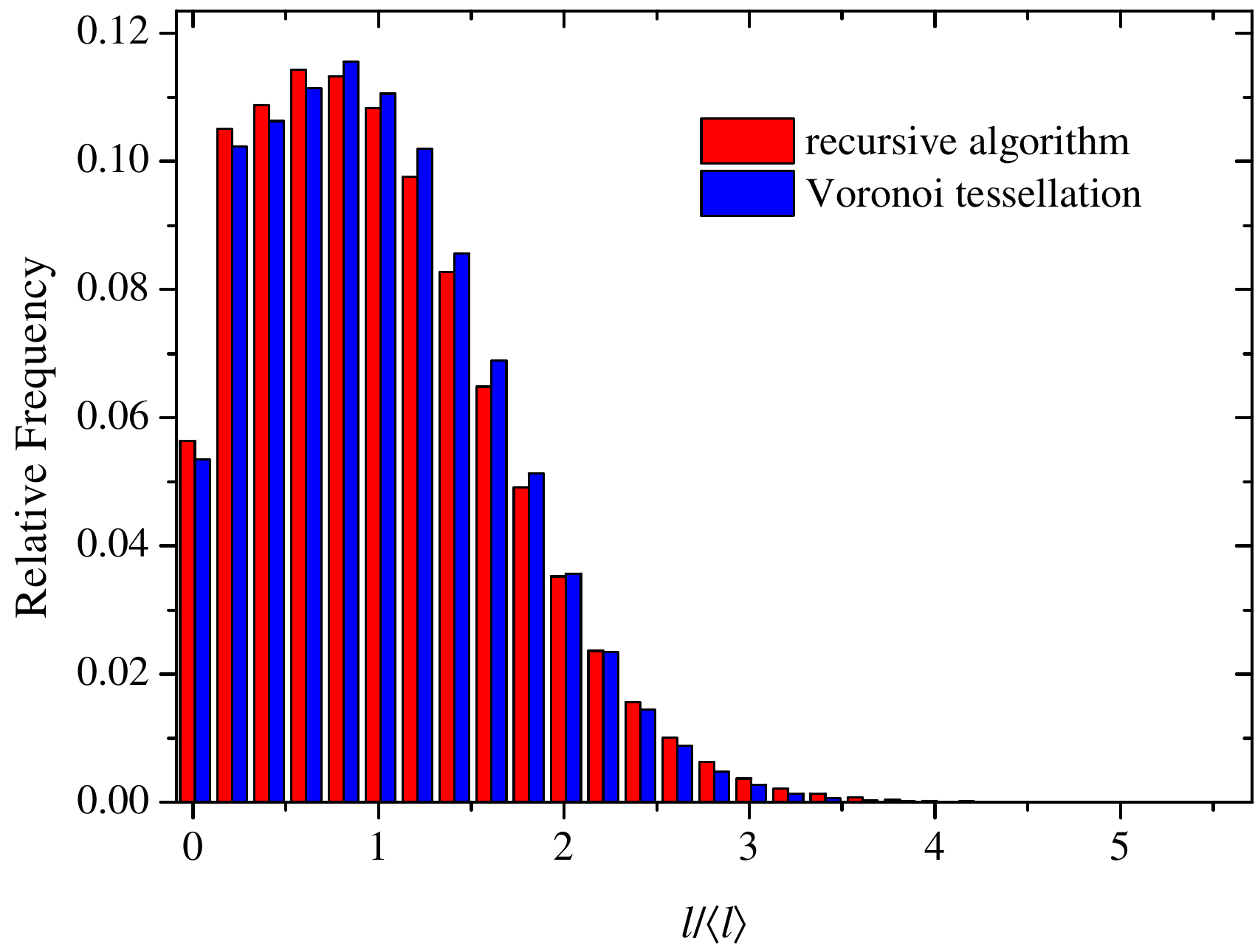}
  \caption{Distribution of edge lengths. Comparison of results obtained using a regular Voronoi tessellation and the recursive algorithm.}\label{fig:length}
\end{figure}

The figure~\ref{fig:length} shows the distribution of edge lengths by size. The resulting distribution is close to that obtained for ordinary Voronoi diagrams~\cite{Tarasevich2023}.

\section{Conclusion}
The use of the recursive algorithm mainly affects the mutual orientation of the edges: if in the case of the Voronoi partition the maximum is approximately at $120^\circ$ ($Y$-shaped crack connections), then in the case of the recursive algorithm the maximum shifts towards $90^\circ$ ($T$-shaped crack connections). The distributions of the remaining characteristics studied differ only in a slight shift of the maximum towards smaller values in the case of the recursive algorithm compared to the usual Voronoi partition.

In case of using the recursive algorithm, the relative length of cracks of different orders is given in Table~\ref{tab:relleng}. The weighted average width of the conductors is $0.71$.
\begin{table}[!htb]
  \caption{Relative length of cracks of different orders}\label{tab:relleng}
\begin{ruledtabular}
  \begin{tabular}{ccc}
     Crack order & fraction & Width, arb.units\\
          \hline
1 & 0.123 & 1.0\\
2 & 0.300 & 0.8\\
3 & 0.577 & 0.6\\
   \end{tabular}
   \end{ruledtabular}
\end{table}

Conductivity calculations can be conveniently compared using the reduced conductivity
$$
\frac{G}{\sigma_0 w h},
$$
where $\sigma_0$ is the conductivity of the material, $w$ and $h$ are the width and thickness of the conductors, respectively.
For a resistance network constructed using the Voronoi diagram, the reduced conductivity is $0.5087\sqrt{n_\text{E}}$~\cite{Tarasevich2023a}. The conductivity value averaged over two directions when using a network constructed using a recursive algorithm, given the weighted average width of the conductors, yields $0.47\sqrt{n_\text{E}}$, where $n_\text{E}$ is the number density of conductive edges.

\begin{acknowledgments}
We acknowledge funding from the Russian Science Foundation, Grant No. 23-21-00074 (I.V.V. and A.V.E.).
\end{acknowledgments}

\end{document}